\documentclass[aps,prl,preprint,superscriptaddress]{revtex4}
\usepackage{graphicx}
\usepackage{amsmath,amssymb}

\begin{document}

\title{Hybrid Exotic Meson Decay Width}
\thanks{This material is based on work supported by the U.S. National Science 
        Foundation under Grant No. PHY-0300065 and upon resources provided by the 
        Lattice Hadron Physics Collaboration through the SciDac program of the 
        U. S. Department of Energy.}

\author{M. S. Cook}
%\email[]{mcook003@fiu.edu}
\author{H. R. Fiebig}
%\email[]{fiebig@fiu.edu}

\affiliation{Department of Physics, Florida International University, \\ 
        Miami, Florida, USA 33199}

\collaboration{LHP Collaboration}

\date{\today}

\begin{abstract}
We present results of a decay width calculation for a hybrid exotic 
meson($h,\,J^{PC}=1^{-+}$) 
in the decay channel $h\rightarrow \pi a_1$. This calculation uses quenched lattice QCD and 
L{\"u}scher's finite box method. Operators for the $h$ and $\pi a_1$ states are used in a 
correlation matrix which was expanded by varying the smearing and fuzzing levels at source 
and sink points.
Scattering phase shifts for a discrete set of relative $\pi a_1$ momenta are determined using 
eigenvalues of 
the correlation matrix and formulae derived by L{\"u}scher.
The phase shift data is very sparse, but fits to Breit-Wigner models are made, resulting
in a decay width of about 80 MeV.
\end{abstract}

%\pacs{}

\maketitle

\section{Introduction}
Most states in the known hadron spectrum are resonances. Their decay widths, arguably more so
than their masses, are sensitive indicators of the intrinsic structure and forces between the constituents.  
The decay of molecule-like states, for example, may be driven by a residual (effective) force and thus
give rise to widths typical of nuclear physics. Spatially more compact hadrons could follow a quite different
mechanism.

Decays can be studied with lattice techniques \cite{Michael:2005}.
In particular, lattice QCD provides us with {\em ab initio} access to unusual hadrons,
like hybrids and exotics, which are a focus of some experimental programs currently under way.
We here give an overview about a quenched lattice simulation set up to estimate
the decay width of a $J^{PC}= 1^{-+}$ hybrid exotic meson resonance, say $h$, coupled to a
$\pi a_1$ decay channel.
Correlation matrices based on the corresponding operators yield (excited) mass spectra.
L{\"u}scher's finite box method \cite{Luscher:1991cf,Luscher:1991ux} is then applied
to get the scattering phase shifts at a discrete set of relative $\pi a_1$ momenta.
A decay width is determined by fitting Breit-Wigner functions. 

Such a simulation is viable, on a reasonably sized lattice, if the energy spectrum exhibits
a level crossing between states
excited by $h$ and $\pi a_1$ operators, preferably, close to the physical value
of the pion mass. Unlike in $\rho\rightarrow \pi\pi$, where this is very hard
to achieve, $h\rightarrow\pi a_1$ allows a relative $s$-wave in the decay channel  
and may thus proceed with relative lattice momentum zero.
The feasibility of the current work much relies on this fact.

\section{Lattice simulation}
Hybrid mesons are quark-antiquark pairs having valence gluons as structural components.
Following \cite{Bernard:1997ex} our choice for the hybrid operator is
\begin{equation}\label{Oh}
{ O}_{h^+;j}(t)=\frac{1}{\sqrt{V}}\sum_{\vec{x}}\sum_{i=1}^3
\bar{d}_{a}(\vec{x}t) \gamma_{i} u_b(\vec{x}t)
(F^{ab}_{ij}(\vec{x}t)-F^{\dagger ab}_{ij}(\vec{x}t))\,.
\end{equation}
Here $a,b$ are color indices and
$F_{ij}$ is a product of SU(3) link matrices defined on paths forming a clover leaf
in the $i$--$j$ plane with center at $\vec{x}$, and $V$ is the spatial lattice volume.
The quantum numbers are $I=1$ and $J^{PC}=1^{-+}$, where the combination $F_{ij}-F_{ij}^\dagger$ is needed
for positive charge conjugation.
Appropriate operators for $\pi a_1$ can be constructed from
\begin{equation}\label{Opia1}
{ O}_{\pi^{+}a_{1}^{0};j}(t)= \frac{1}{\sqrt{V}}\sum_{\vec{x}}\sum_{\vec{y}}
\delta_{\textstyle\vec{y}-\vec{x},\vec{r}}\;
\bar{d}(\vec{x}t)\gamma_{5} u(\vec{x}t) \; 
\bar{d}(\vec{y}t) \gamma_{5} \gamma_{j} d(\vec{y}t)\,.
\end{equation}
The simplest spatial configuration for an $s$-wave ($A_1$ irrep) is the choice $\vec{r}=0$,
which we use in this work.
The elements in the resulting correlation matrix,
\begin{equation}\label{CorrM}
{ C}(t,t_0)=\begin{pmatrix}
{ C}_{h,h}(t,t_0)       & { C}_{h,\pi a_1}(t,t_0) \\
{ C}_{\pi a_1,h}(t,t_0) & { C}_{\pi a_1,\pi a_1}(t,t_0)   \end{pmatrix} \,,
\,\end{equation}
are understood to include spin traces, for example
${ C}_{h,h}(t,t_0)=\sum_{j=1}^3 \langle { O}_{h^+;j}(t) { O}^\dagger_{h^+;j}(t_0)\rangle$,
and so on.

The flavor structure of the operator (\ref{Opia1}) is $\pi a_1\sim\bar{d}u\bar{d}d$,
which gives rise to (two) equal time contractions $\overline{\bar{d}(t)d}(t)$ in (\ref{CorrM}).
To avoid having to compute those propagator elements we observe that these do not occur for
$\pi K_1\sim\bar{d}u\bar{d}s$. Since the experimental $a_1$ and $K_1$ masses are very close, 1260~MeV and
1270~MeV respectively, and given that only the mass spectrum is input to L{\"u}scher's method we argue that
neglecting equal time contractions in ${ C}_{\pi a_1,\pi a_1}$ should not significantly effect our results. 
It should be emphasized that this approximation only is applied to ${ C}_{\pi a_1,\pi a_1}$ in
(\ref{CorrM}), the off-diagonal correlators can be computed or inferred using hermiticity. 

Wuppertal style smearing is done on the quark fields, using APE style gauge link fuzzing in the
process. We use a common strength factor ($\alpha=2.5$) and either $1,2,$ and $3$ smearing iterations.
The same smearing is done at source and sink points.
In this way the correlator (\ref{CorrM}) expands to a $6\times6$ hermitian matrix
\begin{equation}\label{CorrMS}
C(t,t_0)_{2\times 2}\longrightarrow{\mathcal C}(t,t_0)_{6\times 6}\,.
\end{equation}

\begin{figure}%\label{effa1pionhyb}
\centerline{
\includegraphics[width=65mm,height=90mm,angle=90]{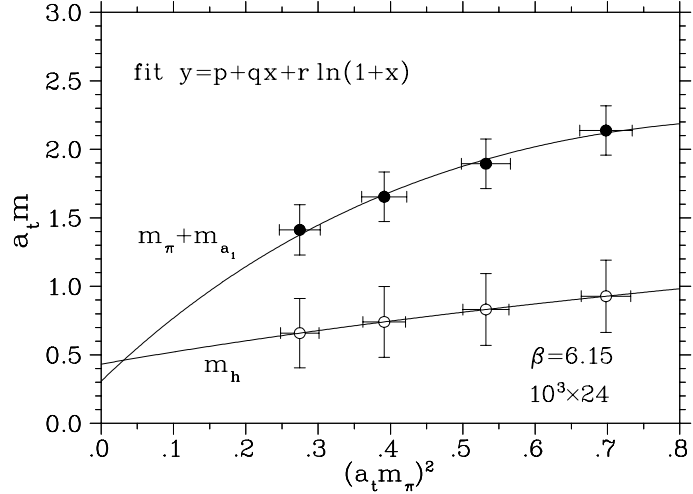} }
%/export/bayes1/mcook/tmp3/effect_a1pion_hybridmass_vs_pionsq.pcm
\caption{Masses of the $h$ and $\pi+a_{1}$ non-interacting system versus the pion mass squared.
The extrapolation shows an energy level crossing at light pion mass.
If the physical scale is set to the $a_1$ meson mass, $1.26$ GeV, then
the hybrid mass extrapolates to $1.73\,\pm\,.39$ GeV. If the $\rho$ meson sets the scale, then
the hybrid mass extrapolates to $1.46\,\pm\,.31$ GeV. }
\label{effa1pionhyb}
\end{figure}

Simulations are made using the anisotropic Wilson gauge action with bare aspect ratio $a_s/a_t=2$
on lattice sizes $L^3\times 24,\: L=8,10,12$, at $\beta=6.15$, and the corresponding Wilson fermion action
in quenched approximation. Four hopping parameters, $\kappa=0.140,\, 0.136,\, 0.132,\, 0.128$,
were used with a multiple mass solver~\cite{Glassner:1996gz}.
Figure.~\ref{effa1pionhyb} shows the ground state masses of the $h$ and the $\pi a_1$ systems versus $m_\pi^2$.
These are obtained using effective mass functions based on output of individual $h$, $\pi$, and $a_1$ correlators
at smearing level two.
The three parameter fit model, see inset in Fig.~\ref{effa1pionhyb}, matches most of the energy-versus-$m_\pi^2$ data
produced in this work. Its choice is purely empirical.
Setting the physical scale to the $\rho$ meson mass  results in
a lattice constant of $a_t=0.07$ fm, and this value is used consistently for the width calculations.

\section{Analysis}
\begin{figure}
\centerline{
\includegraphics[width=65mm,height=90mm,angle=90]{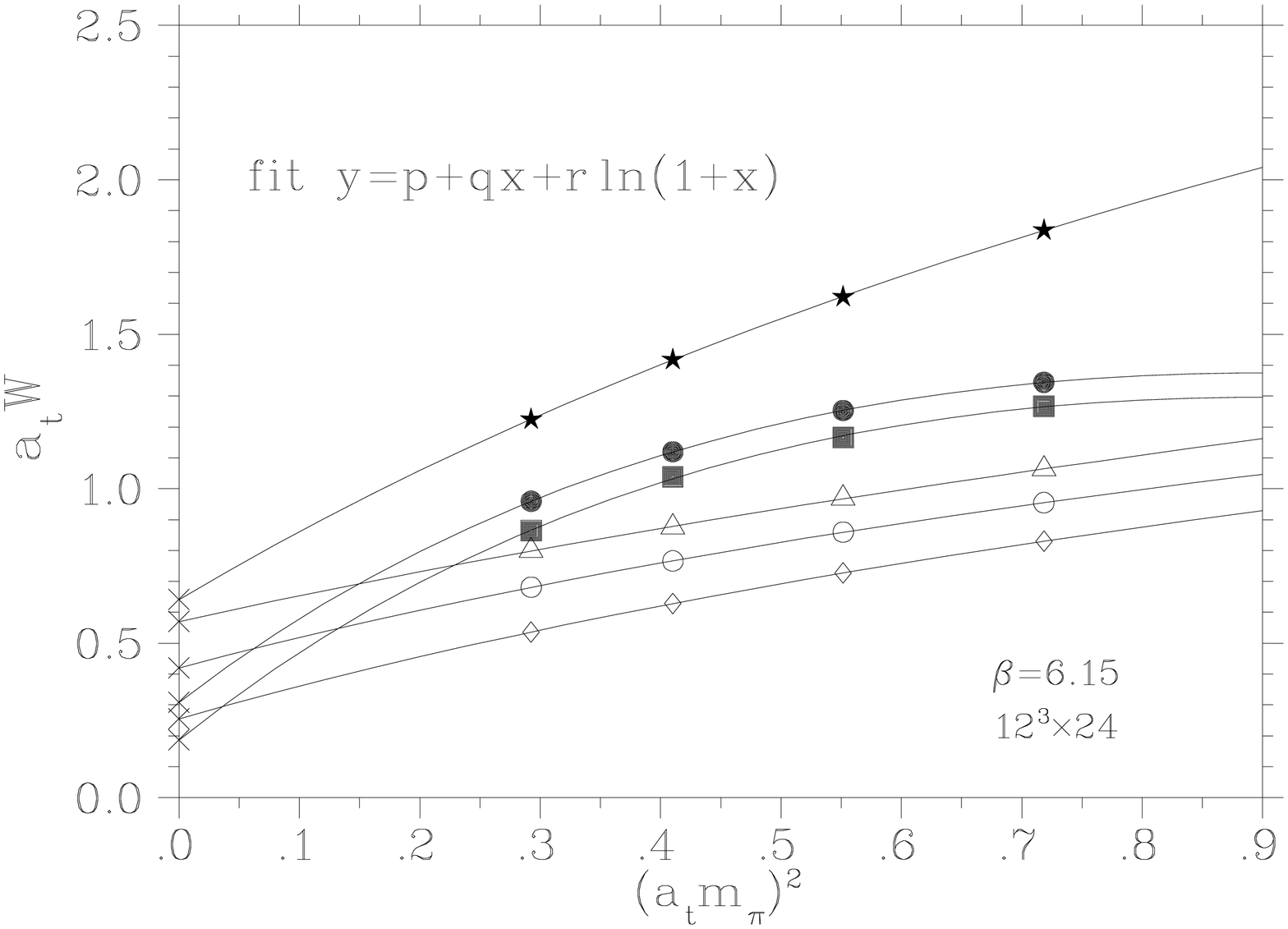}}
%/export/bayes1/mcook/tmp8/effect_lamdaall_log_k_q_mpionsq.pcm
\vspace{8pt}
\centerline{
\includegraphics[width=65mm,height=90mm,angle=90]{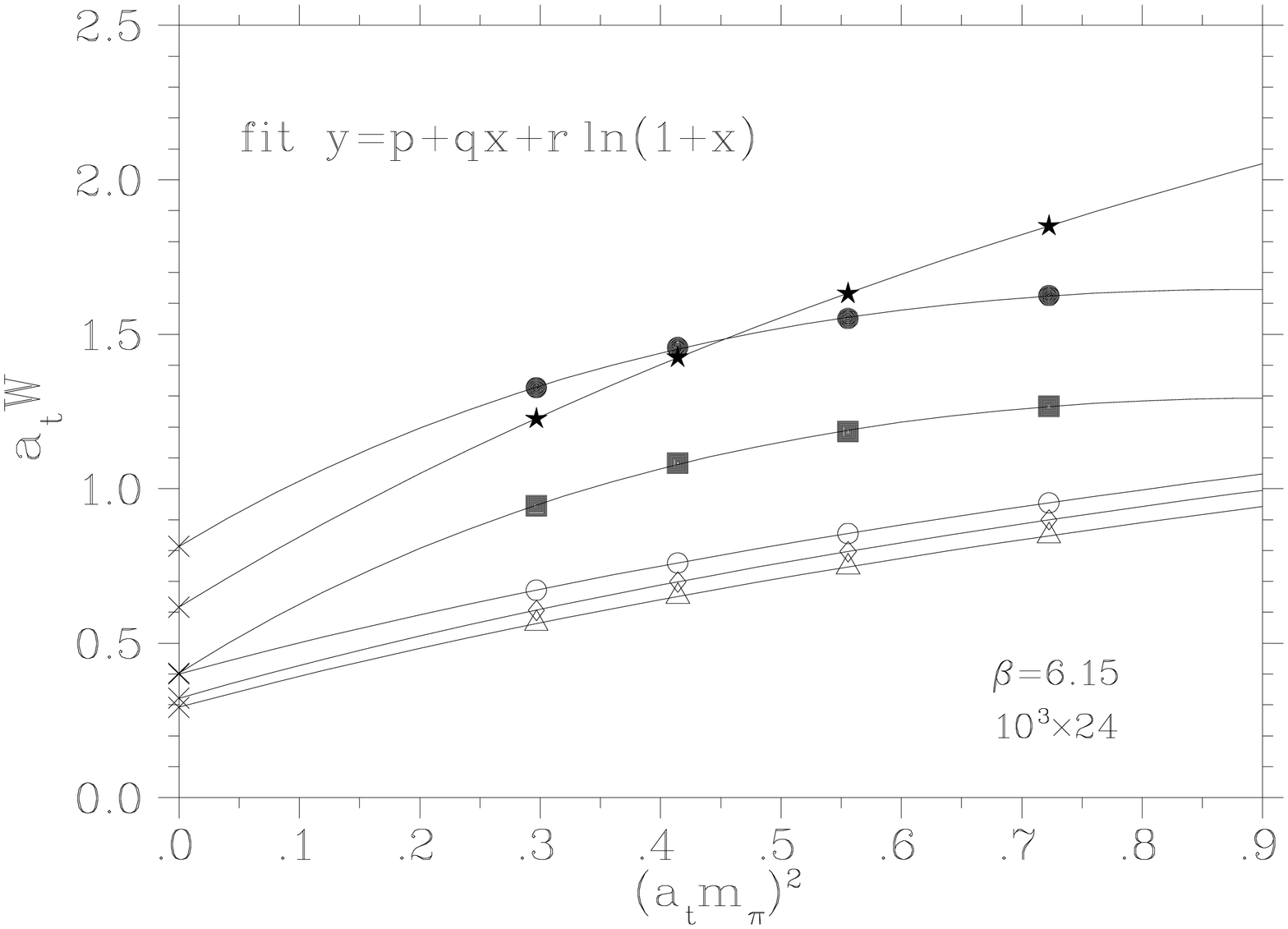}}
%/export/bayes1/mcook/tmp3/effect_lamdaall_log_k_q_mpionsq.pcm 
\caption{Energy spectra $W_n$, $n=1\ldots 6$, and extrapolation to small pion mass for the
$12^3\times24$ and $10^3\times24$ lattices. Error bars are omitted for clarity.}
\label{spectrum1210}
\end{figure}

To extract the mass spectrum the $6\times 6$ correlation matrix (\ref{CorrMS}) is first
diagonalized on a fixed timeslice $t_1$,
\begin{equation} 
{\mathcal C}(t_1,t_0) = V(t_1,t_0) D(t_1,t_0) V^{\dag}(t_1,t_0)\,, 
\end{equation}
where $V(t_1,t_0)$ is unitary and $D(t_1,t_0)$ is diagonal and positive definite.
To ensure the latter $t_1$ should be an {`}early{'} timeslice, we here use $t_1-t_0=4$.
The correlation matrix is then subject to a basis transformation and normalization according to
\begin{equation}\label{Cdiag}
%\widetilde{\cal C}(t)=V(5)\:\cdot\: {\cal C}_{n}(t)\:\cdot\:V^{\dag}(5)
\Lambda(t) = \frac{1}{\sqrt{D(t_1,t_0)}}
V^{\dag}(t_1,t_0) {\mathcal C}(t,t_0) V(t_1,t_0) \frac{1}{\sqrt{D(t_1,t_0)}}\,.
\end{equation}
This procedure amounts to solving the usual generalized eigenvalue problem.
The only modification being that the eigenvectors (columns of $V$) used for the
basis transformation in (\ref{Cdiag}) are employed at a fixed timeslice.
Thus we find that $\Lambda(t)$ is approximately diagonal.
We will simply refer to the diagonal elements, $\Lambda_{nn}(t)$, as the eigenvalues, $\lambda_n(t)$.
Compared to `diagonalizing at all $t$' statistical fluctuations are reduced.
The eigenvalues $\lambda_n(t)$, $n=1\dots 6$, give rise to effective mass functions from
which the spectral energies $W_n$ are obtained.
Energy spectra for the $12^3\times24$ and $10^3\times24$ lattices are shown in Fig.~\ref{spectrum1210}.

\begin{figure}
\centerline{
\includegraphics[width=70mm,height=90mm,angle=90]{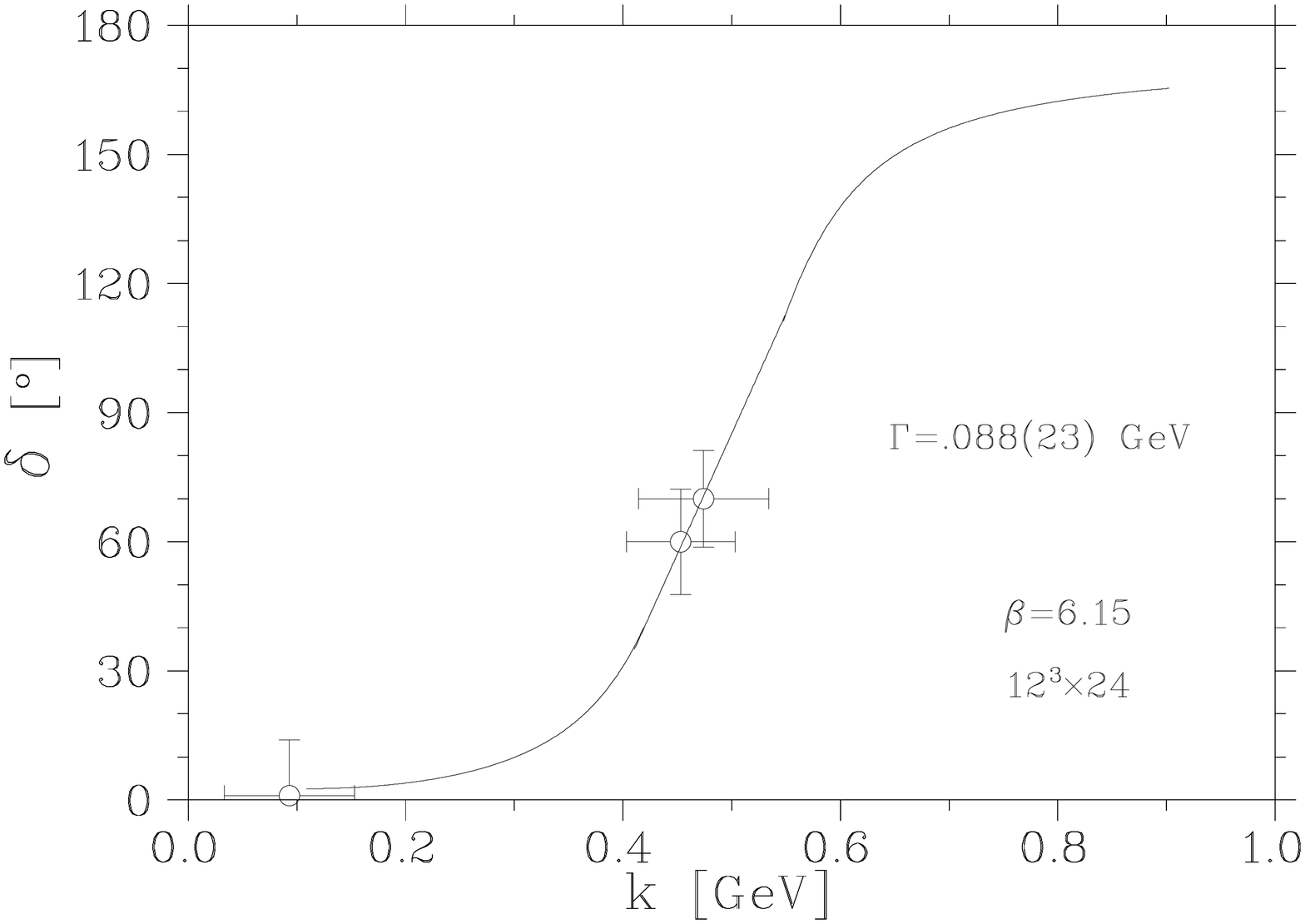}}
%/export/bayes1/mcook/tmp8/phase_momemtum_extrapvalues_12.pcm
\vspace{8pt}
\centerline{
\includegraphics[width=70mm,height=90mm,angle=90]{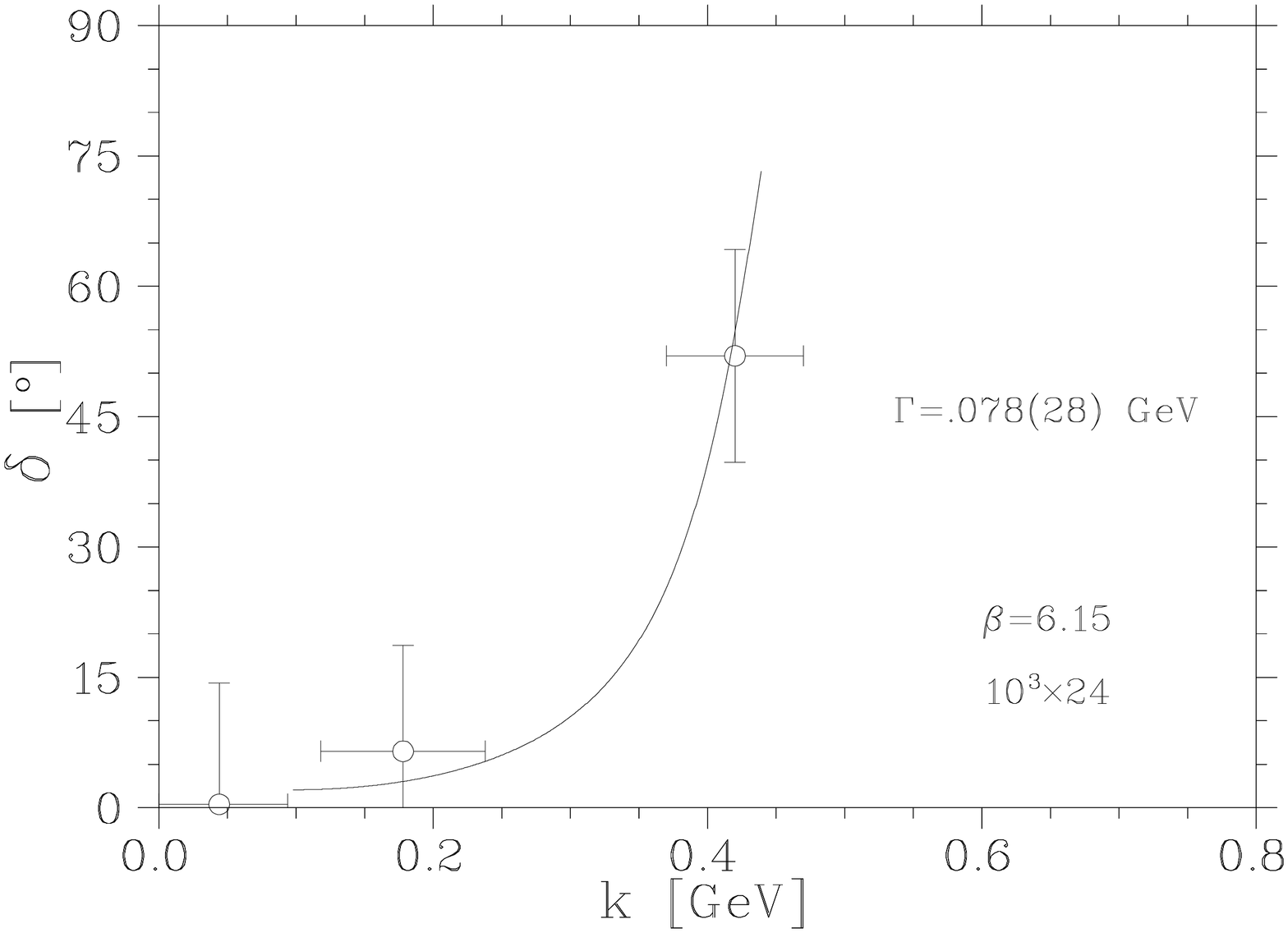}}
%/export/bayes1/mcook/tmp3/phase_momemtum_extrapvalues_10.pcm
\caption {Scattering phase shift versus momentum for the $12^3\times 24$ and $10^3\times 24$ lattices 
using extrapolated
spectra from Fig.~\ref{spectrum1210}. The curve fits are Breit-Wigner models (\protect\ref{BW}), giving
widths of $88\,\pm\,23$ MeV and $78\,\pm\,28$ MeV respectively.}
\label{phase1210}
\end{figure}

With the spectra known, subsets of masses $W_n$ are selected
that fall within the elastic region,
\begin{equation}
(m_{\pi}+m_{a_1})< W_{n} < 2(m_{\pi}+m_{a_1})\,.
\label{Welastic}
\end{equation}
All masses in (\ref{Welastic}) are taken at each of the four values of $m_\pi$
and in the limit $m_\pi\rightarrow 0$  using the extrapolation.
For those sets the relativistic dispersion relation
\begin{equation}
W_n=\sqrt{m_{\pi}^2+k^{2}_{n}}+\sqrt{m_{a_1}^2+k^{2}_{n}}
\end{equation}
is solved for the relative $\pi a_1$ momenta $k_n$. Note, $k_n$ is a prediction for the relative momentum
of a $\pi a_1$ system and is not subject to lattice discretization.
These momenta are input to L{\"u}scher's formula \cite{Luscher:1991cf} for the
s-wave scattering phase shifts $\delta_{n}$,
\begin{equation}
\tan\delta_{n}=-\frac{\pi^{3/2}q_{n}}{{\cal Z}(1;q_{n}^2)} \quad , \quad q_{n}=
\frac{k_{n}L_s}{2\pi}\,.
\end{equation}
Here ${\cal Z}(1;q^2)$ is a generalized $\zeta$-function,
and $L_s=La_s$ is the physical size of the spatial box (using the bare anisotropy, $a_s=2a_t$).
We then attempt fits to all sets of phase shift data with a Breit-Wigner model,
\begin{equation}
\tan\delta(k)=\frac{\Gamma/2}{E_{0}-W(k)}\quad  , \quad \mbox{with}\quad 
W(k)=\sqrt{m_{\pi}^2+k^{2}}+\sqrt{m_{a_1}^2+k^{2}}\quad ,
\label{BW}
\end{equation}
where $E_0$ and $\Gamma$ are parameters. When the fit is successful, $\Gamma$ may be interpreted as a decay width.
In Fig.~\ref{phase1210} we show selected results for
the scattering phase shift data and the corresponding Breit-Wigner fits. It is apparent 
that most of the scattering
phase angle data does not alone justify the use of Breit-Wigner functions. 
Use of these functions requires one 
to assume {\itshape apriori }that a resonant condition exists.

\section{Result}

Physical decay widths, found by using extrapolated energy spectra, varied from 
$78\,\pm\,28$ MeV to $88\,\pm\,23$ MeV for the three lattice volumes tested.
A large portion of the statistical error came from the hybrid correlator. As shown in
Fig.~\ref{phase1210}, the phase shift
data is very sparse. The best fits to the
Breit-Wigner model returned a resonance energy of $E_0 \approx 1.9$ GeV.

Phase shift data obtained
using non-extrapolated spectra did not match up well to the Breit-Wigner model.\, The reason
being, as shown in Fig.~\ref{effa1pionhyb}, the pion masses used are quite far
from the actual energy level crossing, and thus, one is less 
likely to observe resonance behavior.

\section{Summary}

Using  L\"{u}scher's method, a decay width for the $1^{-+}$ exotic meson is 
calculated to be about 80 MeV with a statistical error of 25 MeV.
The number of data points available to fit Breit-Wigner functions is sparse. 
There is a definite energy level crossing between the
$h$ and $\pi a_1$ system at light pion mass giving credence to the simulation.

%\bibliography{latref4}

\end{document}